\documentclass[11pt,twoside]{article}
\usepackage[cp1251]{inputenc}  
\usepackage{graphicx}
\usepackage{amsmath}
\usepackage{amssymb}
\usepackage{mathrsfs}

\begin{document}

\thispagestyle{plain}

\newcommand{\bm}{\boldsymbol}
\newcommand{\mbf}{\mathbf}
\newcommand{\pst}{\hspace*{1.5em}}
\newcommand{\be}{\begin{equation}}
\newcommand{\ee}{\end{equation}}
\newcommand{\ds}{\displaystyle}
\newcommand{\bdm}{\begin{displaymath}}
\newcommand{\edm}{\end{displaymath}}
\newcommand{\bea}{\begin{eqnarray}}
\newcommand{\eea}{\end{eqnarray}}
\newcommand{\rmi}{\mathrm{i}}
\newcommand{\Tr}{\mathrm{Tr}}

\begin{center} {\Large \bf
\begin{tabular}{c}
Partial symplectic quantum tomography schemes. \\
Observables, evolution equations, 
and stationary states equations
\end{tabular}
 } \end{center}

\smallskip

\begin{center} {\bf Ya. A. Korennoy, V. I. Man'ko}\end{center}

\smallskip

\begin{center}
{\it P.N.    Lebedev Physical Institute,                          \\
       Leninskii prospect 53, 119991, Moscow, Russia }
\end{center}

\begin{abstract}\noindent
Partial symplectic conditional and joint probability representations of quantum mechanics
are considered. The correspondence rules for most interesting physical operators are found
and the expressions of the dual symbols of operators are derived.
Calculations were made by use of general formalism 
of quantizers and dequantizers determining the star product
quantization scheme  in these representations. 
Taking the Gaussian functions 
as the distributions of the tomographic parameters 
the examples of joint probability representations were considered. 
Evolution equations and stationary states equations
for partial symplectic conditional and joint probability distributions are obtained.
\end{abstract}


%\vspace{10mm}
%----------------------------------------------------------------------------
\section{Introduction}
%----------------------------------------------------------------------------

Initially the quantum optical tomogram $w(X,\theta)$ was introduced as a tool 
for measuring the quantum state of radiation \cite{BerBer, VogRis}.
Generalizing the optical tomography technique the symplectic tomography
representation of quantum mechanics 
was introduced \cite{Mancini95}, \cite{Mancini96}, \cite{ManciniFoundPhys97} 
(for a review see \cite{IbortPhysScr}).  

According to this representation the states of quantum systems are associated with fair 
probability distributions called quantum tomograms. 
The density operators of the quantum states can be determined from the tomograms,  and
consequently, the tomograms contain the complete information of the quantum properties
equivalent to the information embraced in all of the forms of the density operators
like Wigner function \cite{Wigner32}, Husimi function \cite{Husimi40}, 
Glauber-Sudarshan function \cite{Glauber63, Sudarshan63}.

The problem of dual symbols of physical observables in the symplectic
tomography representation was considered in Ref.~\cite{OlgaJRLR97}.
Evolution equations for optical tomograms of spinless quantum systems
were obtained in Refs.~\cite{KorJRLR3274,KorJRLR32338}.
For the particles with spin it was done in~\cite{KorJRLR36534, KorIntJTP0163112}.
The correspondence rules and dual symbols of operators
in the optical tomography representation were obtained in \cite{KorPhysRevA85}.

The quantum state tomograms depend on extra parameters,
for example, the optical tomogram $w(X,\theta)$ \cite{BerBer, VogRis}
depend on the random position called quadrature component and the
parameter $\theta$ called local oscillator phase. 
It was pointed out 
\cite{BeautyInPhys, PhysScrT153} that the tomogram can be interpreted 
as a conditional probability distribution denoted as $w(X,\theta)\equiv w(X|\theta)$,
and such an interpretation provides the possibility to introduce the joint probability
distribution of two random variables $\widetilde w(X,\theta)$, which determines the optical tomogram
via Bayes' formula \cite{Bayes}.

The symplectic tomogram $M(X,\mu,\nu)$ \cite{Mancini95} represents the distribution function
of the position quadrature $X$ of rotated and squeezed (stretched) phase plane
determined by the parameters $\mu$ and $\nu$.
Thus, $M(X,\mu,\nu)\equiv  M(X|\mu,\nu)$ is a conditional distribution function of the variable 
$X$ under the condition of given $(\mu,\nu)$, and if the  distribution function for 
$\mu$ and $\nu$ is known, we can introduce the joint probability distribution 
$\widetilde M(X,\mu,\nu)$ of three random variables \cite{BeautyInPhys, PhysScrT153}.
Other tomographic schemes (like, e.g., spin tomography) also enable to construct
joint probability representations with all random variables (indices). 

Unfortunately, symplectic tomograms contain an excessive number of variables and parameters.
Our aim in this paper is to introduce partial symplectic conditional and joint probability 
distribution functions without redundant variables.
Also aim of this paper is derivation of correspondence rules and 
symbols of operators in the partial symplectic conditional and  joint probability representation;
and foundation of evolution equations and stationary states equations for such probability
distributions. 

The paper is organized as follows. 
In Section 2 we introduce the  partial symplectic conditional probability representations of states of quantum 
systems using the general formalism of quantizers and dequantizers.
In Section 3 we discuss correspondence rules for the operators, symbols of operators,
the evolution equation, and the equation of stationary states  in such representations.
In Section 4 we consider partial symplectic joint probability representations 
in the $N-$dimensional case and a specific example of  representation with Gaussian distribution functions 
for the tomographic parameters $\bm\mu$ and $\bm\nu$.
Conclusions are given in Section 5.

%----------------------------------------------------------------------------
\section{Partial symplectic (conditional) probability representations of states of quantum systems}
%----------------------------------------------------------------------------
Let  the density matrix $\hat\rho(t)$ dependent on time $t$ and normalized by the condition
$\mathrm{Tr}\{\hat\rho(t)\}=1$ corresponds to the state of the quantum system,
then in the tomographic representation this state is described by the tomographic distribution function
$\mathcal F(x,\eta,t)$ normalized by the condition
\be			\label{Normgeneral}
\int\mathcal F(x,\eta,t)dx=1,
\ee
where $x$  is a set of distribution variables and $\eta$ is a set of parameters 
of corresponding tomography.
According to the star product scheme (see \cite{SIGMA10086}), the tomogram is associated
with the density matrix in the following way:
\be			\label{DefTomgeneral}
\mathcal F(x,\eta,t)=\mathrm{Tr}\left\{
\hat\rho(t)\hat U_{\mathcal F}(x,\eta)
\right\},~~~
\hat\rho(t)=\int \hat D_{\mathcal F}(x,\eta)
\mathcal F(x,\eta,t) dxd\eta,
\ee
where $\hat U_{\mathcal F}(x,\eta)$ and $\hat D_{\mathcal F}(x,\eta)$ are
dequantizer and quantizer operators for appropriate tomographic scheme.

If we have spinless quantum system in the $N-$dimensional space,
then dequantizer and quantizer for the symplectic tomography \cite{OVMankoJPhysA2002} can be written as:
\be		\label{dequantizerSYMP}
\hat U_M(\mbf X,\bm\mu,\bm\nu)=|\mbf X,\bm\mu,\bm\nu\,\rangle\langle \mbf X,\bm\mu,\bm\nu\,|=
\prod_{\sigma=1}^N
\delta(X_\sigma-\hat q_\sigma\mu_\sigma-\hat p_\sigma\nu_\sigma),
\ee
\be		\label{quantizerSYMP}
\hat D_M(\mbf X,\bm\mu,\bm\nu)=
\prod_{\sigma=1}^N\frac{m_\sigma\omega_{\sigma}}{2\pi}
\exp\left\{i\sqrt{\frac{m_\sigma\omega_{\sigma}}{\hbar}}
\left(X_\sigma-\hat q_\sigma\mu_\sigma-\hat p_\sigma\nu_\sigma\right)\right\},
\ee
where $|\mbf X,\bm\mu,\bm\nu\,\rangle$ is an eigenfunction of the operator
$\hat{\mbf X}(\bm\mu,\bm\nu)$ with components
$\hat X_\sigma=\mu_\sigma\hat q_\sigma+\nu_\sigma\hat p_\sigma$
corresponding to the eigenvalue $\mbf X$.

Symplectic tomogram can also be found from the Wigner function
$W(\mbf q,\mbf p,t)$:
\be			\label{RelSympWig}
M(\mbf X,\bm\mu,\bm\nu,t)=\int\prod_{\sigma=1}^N
\delta(X_\sigma-\mu_\sigma q_\sigma-\nu_\sigma p_\sigma)W(\mbf q,\mbf p,t)
d^Nqd^Np,
\ee
which, in turn, is determined by the density matrix in the position representation 
by the well-known formula
\be			\label{Def Wig}
W(\mbf q,\mbf p,t)=\frac{1}{(2\pi\hbar)^N}\int \rho(\mbf q+\mbf u/2,\mbf q-\mbf u/2,t)
e^{-i\mbf p\mbf u/\hbar}d^Nu.
\ee

Symplectic tomogram is redundant, but the redundancy can be extracted.
Consider two following partial symplectic tomographic schemes, in which 
tomograms are defined as:
\be			\label{RelSymp1Wig}
M_1(\mbf X,\bm\nu,t)=\int\prod_{\sigma=1}^N
\delta(X_\sigma-q_\sigma-\nu_\sigma p_\sigma)W(\mbf q,\mbf p,t)
d^Nqd^Np,
\ee
\be			\label{RelSymp2Wig}
M_2(\mbf X,\bm\mu,t)=\int\prod_{\sigma=1}^N
\delta(X_\sigma-\mu_\sigma q_\sigma- p_\sigma)W(\mbf q,\mbf p,t)
d^Nqd^Np.
\ee
These tomograms are obtained from symplectic tomograms when choosing
tomographic parameters, as ${\bm\mu}=1,$ $\bm\nu=1,$ respectively, and they are not redundant.
(Here and below, the dimensional constants $\hbar,~m_\sigma,~\omega_\sigma$ are assumed to be equal to unity.)

Inverse transforms of formulae (\ref{RelSymp1Wig}) and (\ref{RelSymp2Wig})
are given by:
\be			\label{WigSymp1}
W(\mbf q,\mbf p,t)=\frac{1}{(2\pi)^{2N}}\int M_1(\mbf X,\bm\nu,t)\prod_{\sigma=1}^N
|r_\sigma|\exp[ir(X_\sigma-q_\sigma-\nu_\sigma p_\sigma)]
d^NXd^N\nu,
\ee
\be			\label{WigSymp2}
W(\mbf q,\mbf p,t)=\frac{1}{(2\pi)^{2N}}\int M_2(\mbf X,\bm\mu,t)\prod_{\sigma=1}^N
|r_\sigma|\exp[ir(X_\sigma-\mu_\sigma q_\sigma- p_\sigma)]
d^NXd^N\mu.
\ee
Dequantizers and quantizers for these tomograms are expressed with the equations:
\be		\label{dequantizerSYMP1}
\hat U_{M_1}(\mbf X,\bm\nu)=|\mbf X,\bm\nu,1\rangle\langle\mbf X,\bm\nu,1|=
\prod_{\sigma=1}^N
\delta(X_\sigma-\hat q_\sigma-\hat p_\sigma\nu_\sigma),
\ee
\be		\label{quantizerSYMP1}
\hat D_{M_1}(\mbf X,\bm\nu)=
\prod_{\sigma=1}^N\frac{1}{(2\pi)^2}\int |r_\sigma|
\exp\left\{ir_\sigma
\left(X_\sigma-\hat q_\sigma-\hat p_\sigma\nu_\sigma\right)\right\} dr_\sigma,
\ee
\be		\label{dequantizerSYMP2}
\hat U_{M_2}(\mbf X,\bm\mu)=|\mbf X,\bm\mu,2\rangle\langle\mbf X,\bm\mu,2|=
\prod_{\sigma=1}^N
\delta(X_\sigma-\hat q_\sigma\mu_\sigma-\hat p_\sigma),
\ee
\be		\label{quantizerSYMP2}
\hat D_{M_2}(\mbf X,\bm\mu)=
\prod_{\sigma=1}^N\frac{1}{(2\pi)^2}\int |r_\sigma|
\exp\left\{ir_\sigma
\left(X_\sigma-\hat q_\sigma\mu_\sigma-\hat p_\sigma\right)\right\} dr_\sigma,
\ee

where $|\mbf X,\bm\nu,1\rangle$ and $|\mbf X,\bm\mu,2\rangle$ are, respectively, eigenfunctions of the operators
$\hat{\mbf X}_1(\bm\nu)$ and $\hat{\mbf X}_2(\bm\mu)$ with components
$(\hat X_1)_\sigma=\hat q_\sigma+\nu_\sigma\hat p_\sigma$ and 
$(\hat X_2)_\sigma=\mu_\sigma\hat q_\sigma+\hat p_\sigma$
corresponding to the eigenvalue $\mbf X$.

These dequantizers and quantizers satisfy the orthogonality and completeness relations:
\be			\label{ortog1}
\mathrm{Tr}\left\{
\hat U_{M_1}(\mbf X,\bm\nu)\hat D_{M_1}(\mbf X',\bm\nu')
\right\}=\delta(\mbf X-\mbf X')\delta(\bm\nu-\bm\nu'), ~~~
\ee
\be			\label{complete1}
\int
\langle \mbf q_1|\hat U_{M_1}(\mbf X,\bm\nu)|\mbf q_1'\rangle
\langle\mbf q_2|\hat D_{M_1}(\mbf X,\bm\nu)|\mbf q_2'\rangle d^NXd^N\nu
=\delta(\mbf q_1-\mbf q_1')\delta(\mbf q_2-\mbf q_2'), ~~~
\ee
\be			\label{ortog2}
\mathrm{Tr}\left\{
\hat U_{M_2}(\mbf X,\bm\mu)\hat D_{M_2}(\mbf X',\bm\mu')
\right\}=\delta(\mbf X-\mbf X')\delta(\bm\mu-\bm\mu'), ~~~
\ee
\be			\label{complete2}
\int
\langle \mbf q_1|\hat U_{M_2}(\mbf X,\bm\mu)|\mbf q_1'\rangle
\langle\mbf q_2|\hat D_{M_2}(\mbf X,\bm\mu)|\mbf q_2'\rangle d^NXd^N\nu
=\delta(\mbf q_1-\mbf q_1')\delta(\mbf q_2-\mbf q_2').
\ee

%----------------------------------------------------------------------------
\section{Correspondence rules for the operators, evolution equation \\
and stationary states equation}
%----------------------------------------------------------------------------

If an operator $\hat A$ defined on the set of density matrices
$\left\{\hat\rho\right\}$ acts on $\hat\rho$ as $\hat A\hat\rho$,
then, according to the general scheme, the action of this operator on tomogram
${\mathcal F}(x,\eta)$ in the probability representation
can be expressed in terms of dequantizers and quantizers operators $\hat U_{{\mathcal F}},$ 
$\hat D_{{\mathcal F}}$ as follows (for brevity we shall omit the argument $t$,
assuming that the function ${\mathcal F}(x,\eta)$ may depend on time):
\bea			
\big[\hat A\big]_{{\mathcal F}} {\mathcal F}(x,\eta)&=&
\mathrm{Tr}\left\{
\hat U_{{\mathcal F}}(x,\eta)\hat A\int \hat D_{{\mathcal F}}(x',\eta')
{\mathcal F}(x',\eta')dx'd\eta'
\right\}, \nonumber \\[3mm]
&=&
\int\mathrm{Tr}\left\{
\hat U_{{\mathcal F}}(x,\eta)\hat A \hat D_{{\mathcal F}}(x',\eta')
\right\}{\mathcal F}(x',\eta')dx'd\eta',
\label{eq6}
\eea
that is, in this representation the operator $\big[\hat A\big]_{{\mathcal F}}$
is, generally speaking, an integral operator with the kernel
\be			\label{eq7}
\mathcal{K}(x,\eta,x',\eta')=\mathrm{Tr}\left\{
\hat U_{{\mathcal F}}(x,\eta)\hat A \hat D_{{\mathcal F}}(x',\eta')
\right\}.
\ee
With the help of formulae (\ref{eq6}) - (\ref{eq7}) and knowing the expression
for any operator in the density matrix representation
we can find its expression in the probability representation. For position and momentum operators
after calculations we can write:
\be			\label{corrRules1}
\big[\hat q_j\big]_{M_1}=X_j+\nu_j\partial_{\nu_j}\partial_{X_j}^{-1}
+\frac{\rmi\nu_j}{2}\partial_{X_j},~~~~
\big[\hat p_j\big]_{M_1}=-\partial_{\nu_j}\partial_{X_j}^{-1}-\frac{\rmi}{2}\partial_{X_j},
\ee
\be			\label{corrRules2}
\big[\hat q_j\big]_{M_2}=-\partial_{\mu_j}\partial_{X_j}^{-1}+\frac{\rmi}{2}\partial_{X_j},~~~~
\big[\hat p_j\big]_{M_2}=X_j+\mu_j\partial_{\mu_j}\partial_{X_j}^{-1}
-\frac{\rmi\mu_j}{2}\partial_{X_j},
\ee
where we introduced the designation  for inverse derivatives \cite{KorPhysRevA85}
\be			\label{invder}
\partial_{x_\sigma}^{-n}F(x_\sigma)=\frac{1}{(n-1)!}
\int(x_\sigma-x_\sigma')^{n-1}\Theta(x_\sigma-x_\sigma')F(x_\sigma')d x_\sigma',
\ee
where $\Theta(x_\sigma-x_\sigma')$ is a Heaviside step function.

For the sum and for the product of two operators $\hat A$ and $\hat B$
we can write:
\bdm
\big[\hat A+\hat B\big]_{{\mathcal F}}=\big[\hat A\big]_{{\mathcal F}}
+\big[\hat B\big]_{{\mathcal F}},~~~~
\big[\hat A\hat B\big]_{{\mathcal F}}=\big[\hat A\big]_{{\mathcal F}}
\big[\hat B\big]_{{\mathcal F}}\,.
\edm
From these properties it follows that for any analytic function
$R(\hat A_1,\hat A_2,...,\hat A_k)$ on the set of operators
$\hat A_1,\hat A_2,...,\hat A_k$ the equality is fulfilled
\be			\label{ratifuncoper}
\big[R(\hat A_1,\hat A_2,...,\hat A_k)\big]_{{\mathcal F}}=
R\left(\big[\hat A_1\big]_{{\mathcal F}},\big[\hat A_2\big]_{{\mathcal F}},...,
\big[\hat A_k\big]_{{\mathcal F}}\right).
\ee
Thus, in most cases it is sufficient to find the correspondence rules
for position and momentum operators.

For any operator $\hat A$ its symbol ${\mathcal F}_{\hat A}(x,\eta)$ and dual symbol
${\mathcal F}^{(d)}_{\hat A}(x,\eta)$ are also found
in accordance with the general scheme 
using dequantizer and quantizer (\ref{eq5_1})
\be			\label{symbgeneral0}
{\mathcal F}_{\hat A}(x,\eta)=\mathrm{Tr}\left\{\hat A\hat U_{{\mathcal F}}(x,\eta)\right\},
\ee
\be			\label{symbgeneral}
{\mathcal F}^{(d)}_{\hat A}(x,\eta)=\mathrm{Tr}\left\{\hat A\hat D_{{\mathcal F}}(x,\eta)\right\}.
\ee
The average value of the operator $\hat A$ in the state described by the conditional probability distribution
${\mathcal F}(x,\eta)$ is determined by the dual symbol as follows:
\be			\label{averdual}
\langle\hat A\rangle=\int{\mathcal F}^{(d)}_{\hat A}(x,\eta)
{\mathcal F}(x,\eta)dx d\eta.
\ee

It is obvious, that dual symbol for $\hat q_j$ operator in the $M_1(\mbf X,\bm\nu)$ tomographic representation equals
\be			\label{dualsymbq1}
M^{(d)}_{1~\hat q_j}(\mbf X,\bm\nu)=X_j\delta(\nu_j)\delta(\nu_\sigma-\nu_{0\sigma})_{\sigma\neq j},
\ee
where $\nu_{0\sigma}$ are arbitrary real numbers. Taking into account, that
$$
\int X_j\delta(\nu_j-1)M_1(\mbf X,\bm\nu)d^NXd\nu_j=\langle\hat q\rangle+\langle\hat p\rangle,
$$
after calculations we obtain dual symbol for momentum operator $\hat p_j$ 
\be			\label{dualsymbp1}
M^{(d)}_{1~\hat p_j}(\mbf X,\bm\nu)=X_j[\delta(\nu_j-1)-\delta(\nu_j)]\delta(\nu_\sigma-\nu_{0\sigma})_{\sigma\neq j}.
\ee
Symbols of other operators can be calculated analogically.
For example, for $\hat q_j^2,$~ $\hat p_j^2,$~ $\hat p_j\hat q_j+\hat q_j\hat p_j$ we have:
$$
M^{(d)}_{1~\hat q_j^2}(\mbf X,\bm\nu)=X_j^2\delta(\nu_j)\delta(\nu_\sigma-\nu_{0\sigma})_{\sigma\neq j},
$$
$$
M^{(d)}_{1~\hat p_j^2}(\mbf X,\bm\nu)=\frac{X_j^2}{2}[\delta(\nu_j-1)+\delta(\nu_j+1)-2\delta(\nu_j)]
\delta(\nu_\sigma-\nu_{0\sigma})_{\sigma\neq j},
$$
$$
M^{(d)}_{1~\hat p_j\hat q_j+\hat q_j\hat p_j}(\mbf X,\bm\nu)=
\frac{X_j^2}{2}[\delta(\nu_j-1)-\delta(\nu_j+1)]
\delta(\nu_\sigma-\nu_{0\sigma})_{\sigma\neq j}.
$$
In the tomographic representation, in which states are described by the tomogram $M_2(\mbf X,\bm\mu,t),$
similar calculations give rise to the following expressions for dual symbols of these operators:
$$
M^{(d)}_{2~\hat q_j}(\mbf X,\bm\mu)=X_j[\delta(\mu_j-1)-\delta(\mu_j)]\delta(\mu_\sigma-\mu_{0\sigma})_{\sigma\neq j},
$$
$$
M^{(d)}_{2~\hat p_j}(\mbf X,\bm\mu)=X_j\delta(\mu_j)\delta(\mu_\sigma-\mu_{0\sigma})_{\sigma\neq j},
$$
$$
M^{(d)}_{2~\hat q_j^2}(\mbf X,\bm\mu)=\frac{X_j^2}{2}[\delta(\mu_j-1)+\delta(\mu_j+1)-2\delta(\mu_j)]
\delta(\mu_\sigma-\mu_{0\sigma})_{\sigma\neq j},
$$
$$
M^{(d)}_{2~\hat p_j^2}(\mbf X,\bm\mu)=X_j^2\delta(\mu_j)\delta(\mu_\sigma-\mu_{0\sigma})_{\sigma\neq j},
$$
$$
M^{(d)}_{2~\hat p_j\hat q_j+\hat q_j\hat p_j}(\mbf X,\bm\mu)=
\frac{X_j^2}{2}[\delta(\mu_j-1)-\delta(\mu_j+1)]
\delta(\mu_\sigma-\mu_{0\sigma})_{\sigma\neq j}.
$$

The {\bf evolution equation} for the joint probability distribution is found from the von-Neumann equation
\be		\label{vonNeumann}
i\hbar\partial_t\hat\rho=[\hat H,\hat\rho]
\ee
according to the method \cite{KorJRLR32338}
\be			\label{evolvEQ1}
\partial_t{\mathcal F}(x,\eta,t)=\frac{2}{\hbar}\int\mathrm{Im}\left[\mathrm{Tr}
\left\{\hat H(t)\hat D_{{\mathcal F}}(x',\eta')\hat U_{{\mathcal F}}(x,\eta)\right\}\right]{\mathcal F}(x',\eta',t)
dx'd\eta',
\ee
or
\be			\label{evolvEQ2}
\partial_t {\mathcal F}(x,\eta,t)=\frac{2}{\hbar}\mathrm{Im}\hat H
\left([\hat{\mbf q}]_{{\mathcal F}}\,,[\hat{\mbf p}]_{{\mathcal F}},t\right)
{\mathcal F}(x,\eta,t),
\ee
where hereinafter $\partial_t $ is an abbreviated designation of the derivative $\partial/\partial t$.
For the Hamiltonian with potential energy $V(\mbf q,t)$ after calculations we obtain:
\be			\label{evolvEQ2M1}
\partial_t M_1(\mbf X,\bm\nu,t)=\partial_{\bm\nu}M_1(\mbf X,\bm\nu,t)+\frac{2}{\hbar}\mathrm{Im}\hat V
\left(X_j + \nu_j\partial_{\nu_j}\partial_{X_j}^{-1} + \frac{\rmi\nu_j}{2}\partial_{X_j},t\right)
M_1(\mbf X,\bm\nu,t),
\ee
\be			\label{evolvEQ2M2}
\partial_t M_2(\mbf X,\bm\mu,t)=\left[-\sum_{j=1}^N(\mu_j+\mu_jX_j\partial_{X_j}+\mu_j^2\partial_{\mu_j})
+\frac{2}{\hbar}\mathrm{Im}\hat V
\left( - \partial_{\mu_j}\partial_{X_j}^{-1} + \frac{\rmi}{2}\partial_{X_j},t\right)\right]
M_2(\mbf X,\bm\mu,t),
\ee

When the Hamiltonian is time-independent, for the stationary states equation 
\be			\label{statEQ}
\hat H\hat\rho_E=E\hat\rho_E=\hat\rho_E\hat H
\ee
in the probability representation we have:
\be			\label{statEQ1}
E{\mathcal F}_E(x,\eta)=\mathrm{Re}\hat H
\left([\hat{\mbf q}]_{{\mathcal F}}\,,[\hat{\mbf p}]_{{\mathcal F}}\right)
{\mathcal F}_E(x,\eta).
\ee
For conditional probabilities $M_1(\mbf X,\bm\nu),$ $M_2(\mbf X,\bm\nu)$ we have respectively:
\be			\label{statEQ1M1}
EM_1(\mbf X,\bm\nu)=\left[\frac{1}{2}\partial_{\nu_j}^2\partial_{X_j}^{-2}-\frac{1}{8}\partial_{X_j}^2
+\mathrm{Re}\hat V
\left(X_j + \nu_j\partial_{\nu_j}\partial_{X_j}^{-1} + \frac{\rmi\nu_j}{2}\partial_{X_j},t\right)\right]
M_1(\mbf X,\bm\nu,t),
\ee
\be			\label{statEQ1M2}
EM_2(\mbf X,\bm\mu)=\left[\frac{X_j^2}{2}
+X_j\mu_j\partial_{\mu_j}\partial_{X_j}^{-1}
+\frac{\mu_j^2}{2}\partial_{\mu_j}^2\partial_{X_j}^{-2}
-\frac{\mu_j^2}{8}\partial_{X_j}^2
+\mathrm{Re}\hat V
\left( - \partial_{\mu_j}\partial_{X_j}^{-1} + \frac{\rmi}{2}\partial_{X_j},t\right)\right]
M_2(\mbf X,\bm\mu,t).
\ee
Probability distributions ${\mathcal F}_E(x,\eta)$ corresponding to the stationary states 
must also satisfy the stationary condition, which can be written as:
\be			\label{statCond}
\mathrm{Im}\hat H
\left\{[\hat{\mbf q}]_{{\mathcal F}}\,,[\hat{\mbf p}]_{{\mathcal F}}\right\}
{\mathcal F}_E(x,\eta)=0.
\ee

%----------------------------------------------------------------------------
\section{Joint probability representation of states of quantum systems}
%----------------------------------------------------------------------------

Following by Refs. \cite{BeautyInPhys, PhysScrT153} let we introduce the joint probability
distribution function describing the state of a physical system.
For this aim we note that if the set of parameters $\eta$ is chosen randomly with a distribution function
$P(\eta)$ normalized by the condition
\be			\label{normir}
\int P(\eta)d\eta=1,
\ee
then, according to Bayes' formula \cite{Bayes}, dependent (generally speaking) on time
joint probability distribution $\widetilde{\mathcal F}(x,\eta,t)$ of the two sets 
of random variables $X$ and $\eta$ will be
equal
\be			\label{eq1}
\widetilde{\mathcal F}(x,\eta,t)= {\mathcal F}(x,\eta,t)P(\eta).
\ee

Due to normalization property of the tomogram (\ref{Normgeneral})
and normalized distribution function $P(\eta)$ of the set of parameters $\eta$,
the joint probability distribution (\ref{eq1}) will be automatically
normalized, but in the space of two sets of variables $x$ and $\eta$ 
\be			\label{eq2}
\int \widetilde {\mathcal F}(x,\eta,t) dxd\eta=1.
\ee
Further let us make use of the universal star product scheme (see, e.g. \cite{SIGMA10086}).
For this we introduce the corresponding dequantizer and quantizer operators relating the function
$\widetilde{\mathcal F}(x,\eta,t)$ and the density matrix $\hat\rho(t)$. It is evident that
\be			\label{eq3}
\widetilde{\mathcal F}(x,\eta,t) =\mathrm{Tr}\left\{
\hat\rho(t)\hat U_{\mathcal F}(x,\eta)P(\eta)
\right\}=
\mathrm{Tr}\left\{
\hat\rho(t) \hat U_{\widetilde{\mathcal F}}(x,\eta)
\right\},
\ee
\be			\label{eq4}
\hat\rho(t)=\int \hat D_{\mathcal F}(x,\eta)P^{-1}(\eta)
\widetilde{\mathcal F}(x,\eta,t) dxd\eta=
\int\hat D_{\widetilde{\mathcal F}}(x,\eta)
\widetilde{\mathcal F}(x,\eta,t) dxd\eta,
\ee
where $\hat U_{\widetilde{\mathcal F}}(x,\eta)$, $\hat D_{\widetilde{\mathcal F}}(x,\eta)$ are the new
dequantizer and quantizer for the star product scheme in the joint probability representation
\be			\label{eq5_1}
\hat U_{\widetilde{\mathcal F}}(x,\eta)=P(\eta) \hat U_{\mathcal F}(x,\eta),~~~~
\hat D_{\widetilde{\mathcal F}}(x,\eta)=P^{-1}(\eta) \hat D_{\mathcal F}(x,\eta).
\ee

Additionally, in the definition of the partial symplectic joint probability distribution we will assume 
that the distribution functions $P(\bm\nu),$ $P(\bm\mu)$ of  tomographic parameters tend to zero 
at infinity with all of theyr derivatives, and they are integrable across the hyperspaces $\{\bm\nu\},$ $\{\bm\mu\}$
with any finite products of its arguments.

Because of the simple relation between dequantizers and quantizers for the tomographic
representation and for the joint probability representation (\ref{eq5_1}),
the correspondence rules can be found directly from the appropriate rules in the relevant 
tomographic representation. That is, if $\big[\hat A\big]_{\mathcal F}$ is the expression
for the operator $\hat A$ in the tomographic representation, then in the joint probability
representation we, obviously, have:
\be			\label{eq8}
\big[\hat A\big]_{\widetilde {\mathcal F}}=P(\eta)\big[\hat A\big]_{\mathcal F}
P^{-1}(\eta).
\ee
Thus, for example, for the position operator
\be			\label{eq9}
[\hat{\mbf q}]_{\widetilde{\mathcal F}}=P(\eta)[\hat{\mbf q}]_{\mathcal F}
P^{-1}(\eta).
\ee

%____________________________________________
%%%%%%%%%%%%%%%%%%%%%%%%%%%%%
%----------------------------------------------------------------------

{\bf As examples}, consider the cases with the distribution of the tomographic parameters
$\bm\mu$ and $\bm\nu$ in the form of shifted and deformed Gaussian functions
for $N-$dimensional quantum system
\be			\label{shiftGauss}
P_1(\bm{\nu})=\pi^{-N/2}\prod_{\sigma=1}^N \zeta_\sigma^{-1}
\exp\left[-\frac{(\nu_\sigma-\nu_{0\sigma})^2}{\zeta_\sigma^2}\right].
\ee
$$
P_2(\bm{\mu})=\pi^{-N/2}\prod_{\sigma=1}^N\xi_\sigma^{-1} 
\exp\left[-\frac{(\mu_\sigma-\mu_{0\sigma})^2}{\xi_\sigma^2}\right].
$$

From the correspondence rules for the operators of components of position and momentum
in the partial symplectic tomography representation (\ref{corrRules1}),
with the help of formula (\ref{eq8})  we obtain the operators 
of position and momentum in the joint probability representation 
$\widetilde M_1(\mbf X,\bm\nu,t)=P_1(\bm{\nu})M_1(\mbf X,\bm\nu,t)$
\be			\label{QreprP1}
\left[\hat q_j\right]_{\widetilde M_1}=P_1(\bm{\nu})\left(X_j
+\nu_j\partial_{\nu_j}\partial_{X_j}^{-1}
+\frac{\rmi\nu_j}{2}\partial_{X_j} 
\right)P_1^{-1}(\bm{\nu})=
X_j+\nu_j\left(2\frac{\nu_j-\nu_{0j}}{\zeta_j^2}+\partial_{\nu_j}\right)\partial_{X_j}^{-1}
+\frac{\rmi\nu_j}{2}\partial_{X_j},
\ee
\be			\label{PreprP1}
\left[\hat p_j\right]_{\widetilde M_1}=
-\left(2\frac{\nu_j-\nu_{0j}}{\zeta_j^2}+\partial_{\nu_j}\right)\partial_{X_j}^{-1}
-\frac{\rmi}{2}\partial_{X_j}.
\ee
Correspondence rules in the joint probability representation 
$\widetilde M_2(\mbf X,\bm\mu,t)=P_2(\bm{\mu})M_2(\mbf X,\bm\mu,t)$ are derived from
correspondence rules (\ref{corrRules2}):
\be			\label{QreprP2}
\left[\hat q_j\right]_{\widetilde M_2}=
-\left(2\frac{\mu_j-\mu_{0j}}{\xi_j^2}+\partial_{\mu_j}\right)\partial_{X_j}^{-1}
+\frac{\rmi}{2}\partial_{X_j}.
\ee
\be			\label{PreprP2}
\left[\hat p_j\right]_{\widetilde M_2}=
X_j+\mu_j\left(2\frac{\mu_j-\mu_{0j}}{\xi_j^2}+\partial_{\mu_j}\right)\partial_{X_j}^{-1}
-\frac{\rmi\mu_j}{2}\partial_{X_j}.
\ee

Dual symbols of operators in joint probability representations $\widetilde M_1(\mbf X,\bm\nu,t)$
and $\widetilde M_2(\mbf X,\bm\mu,t)$ are calculated using methods similar to that described in our article
\cite{KorIntJTP0163261}. Moreover, such symbols can be expressed in terms of both singular and regular
generalized functions.

Thus, for singular generalized functions:
$$
\widetilde M_{1~\hat q_j}^{(d)}(\mbf X,\bm\nu)=\pi^{1/2}\zeta_j X_j\delta(\nu_j)\exp(\nu_{0j}^2/\zeta_j^2)
\delta(\nu_\sigma-\nu_{0\sigma})_{\sigma\neq j},
$$
$$
\widetilde M_{1~\hat p_j}^{(d)}(\mbf X,\bm\nu)=\pi^{1/2}\zeta_j X_j\Big[
\delta(\nu_j-1)\exp\left((1-\nu_{0j})^2/\zeta_j^2\right)
-\delta(\nu_j)\exp\left(\nu_{0j}^2/\zeta_j^2\right)
\Big]
\delta(\nu_\sigma-\nu_{0\sigma})_{\sigma\neq j},
$$

$$
\widetilde M_{2~\hat q_j}^{(d)}(\mbf X,\bm\mu)=
\pi^{1/2}\xi_j X_j\Big[
\delta(\mu_j-1)\exp\left((1-\mu_{0j})^2/\xi_j^2\right)
-\delta(\mu_j)\exp\left(\mu_{0j}^2/\xi_j^2\right)
\Big]
\delta(\mu_\sigma-\mu_{0\sigma})_{\sigma\neq j},
$$
$$
\widetilde M_{2~\hat p_j}^{(d)}(\mbf X,\bm\mu)=
\pi^{1/2}\xi_j X_j\delta(\mu_j)\exp(\mu_{0j}^2/\xi_j^2)
\delta(\mu_\sigma-\mu_{0\sigma})_{\sigma\neq j}.
$$
Calculation of dual symbols in terms of regular generalized functions leads to the following rezult:
$$
\widetilde M_{1~\hat q}^{(d)}(X,\nu)=
X-2\nu_0\frac{\nu-\nu_0}{\zeta^2}X,
$$
$$
\widetilde M_{1~\hat p}^{(d)}(X,\nu)=
2\frac{\nu-\nu_0}{\zeta^2}X,
$$
$$
\widetilde M_{2~\hat q}^{(d)}(X,\mu)=
2\frac{\mu-\mu_0}{\xi^2}X,
$$
$$
\widetilde M_{2~\hat p}^{(d)}(X,\mu)=
X-2\mu_0\frac{\mu-\mu_0}{\xi^2}X,
$$
(for brevity we present these formulas in one-dimensional case.)
 \\

{\bf Evolution equations and stationary states equations} in representations
of joint distribution functions $\widetilde M_1(\mbf X,\bm\nu,t)$
and $\widetilde M_2(\mbf X,\bm\mu,t)$ are obtained from equations (\ref{evolvEQ2}) 
and (\ref{statEQ1})
using substitutions of correspondence rules (\ref{QreprP1},\ref{PreprP1},\ref{QreprP2},\ref{PreprP2})
\be			\label{evolvEQjoint1}
\partial_t \widetilde M_1(\mbf X,\bm\nu,t)=\frac{2}{\hbar}\mathrm{Im}\hat H
\left([\hat{\mbf q}]_{\widetilde M_1}\,,[\hat{\mbf p}]_{\widetilde M_1},t\right)
\widetilde M_1(\mbf X,\bm\nu,t),
\ee
\be			\label{evolvEQjoint2}
\partial_t \widetilde M_2(\mbf X,\bm\mu,t)=\frac{2}{\hbar}\mathrm{Im}\hat H
\left([\hat{\mbf q}]_{\widetilde M_2}\,,[\hat{\mbf p}]_{\widetilde M_2},t\right)
\widetilde M_2(\mbf X,\bm\mu,t),
\ee

\be			\label{statEQjoint1}
E\widetilde M_1(\mbf X,\bm\nu)=\mathrm{Re}\hat H
\left([\hat{\mbf q}]_{\widetilde M_1}\,,[\hat{\mbf p}]_{\widetilde M_1}\right)
\widetilde M_1(\mbf X,\bm\nu),
\ee
\be			\label{statEQjoint2}
E\widetilde M_2(\mbf X,\bm\mu)=\mathrm{Re}\hat H
\left([\hat{\mbf q}]_{\widetilde M_2}\,,[\hat{\mbf p}]_{\widetilde M_2}\right)
\widetilde M_2(\mbf X,\bm\mu).
\ee

\section{Conclusion}
To summarize, we point out the main results of this work.
We extracted the redundancy of the symplectic tomogram by introducing
partial symplectic tomograms, which are conditional probability distribution 
functions without redundancy.
Based on these partial symplectic tomograms, we introduced joint probability 
distribution functions in which the tomographic parameters are random variables 
with predetermined distribution functions.
The conditional and joint distribution functions we introduced contain as much 
information about a quantum state as the density matrix of this state.
These distribution functions give rise to new representations of quantum mechanics, 
for which we have calculated correspondence rules for operators, symbols of operators, 
evolution equations and stationary state equations for arbitrary Hamiltonians.
The main advantage of the representations we introduced is the non-negativity of distribution 
functions describing quantum states, without redundancy of observed variables 
and tomography parameters.

%\vspace{-8mm}

\end{document}